\documentclass[preprint,showpacs,preprintnumbers,amsmath,amssymb]{revtex4}

\usepackage{graphicx}
\usepackage{dcolumn}
\usepackage{bm}

\begin{document}
\def\bea{\begin{eqnarray}}
\def\eea{\end{eqnarray}}
\def\nn{\nonumber}

\renewcommand\epsilon{\varepsilon}
\def\beq{\begin{equation}}
\def\eeq{\end{equation}}
\def\lla{\left\langle}
\def\rra{\right\rangle}
\def\za{\alpha}
\def\zb{\beta}
\def\lsim{\mathrel{\raise.3ex\hbox{$<$\kern-.75em\lower1ex\h
box{$\sim$}}} }
\def\gsim{\mathrel{\raise.3ex\hbox{$>$\kern-.75em\lower1ex\h
box{$\sim$}}} }
\newcommand{\Rbs}{\mbox{${{\scriptstyle
\not}{\scriptscriptstyle R}}$}}
\newcommand{\cm}{\check{m}}


\title{Discrete symmetries and neutrino masses}

\author{Kim ~Siyeon} \email{siyeon@cau.ac.kr}

\affiliation{Department of Physics,
        Chung-Ang University, Seoul 156-756, Korea}
\date{January 3, 2005}

\begin{abstract}
\noindent We constructed a model of neutrino masses using
Froggatt-Nielsen mechanism with $U(1) \times Z_3 \times Z_2$
flavor symmetry. The model predicts that $(2/3)m_2/m_3 \sim
\sqrt{2}\sin\theta_{13}$ at lepton number violating scale $M_1$.
It is shown that the small values for $m_2/m_3$ and
$\sin\theta_{13}$ are consequences of breaking discrete
symmetries.
\end{abstract}


\maketitle \thispagestyle{empty}

\newcommand{\yuk}{\mathcal{Y}}
\newcommand{\be}{\begin{equation}}
\newcommand{\ee}{\end{equation}}


\section{Introduction}
\noindent Having experimental data at our hands, it is undeniable
that two of three mixing angles in the lepton sector  are large
and the other is small. A few detailed types of mixing angle sets
depending on theories to derive them are still debatable.
Interpreting the atmospheric and solar neutrino experiments
\cite{atm}\cite{SNO} in terms of two-flavor mixing, the mixing
angle for the oscillation of atmospheric neutrinos is understood
to be maximal or nearly maximal: $ \sin^22\theta_{atm} \simeq 1$,
whereas the one for the oscillation of solar neutrinos is not
maximal but large: $ \sin^2\theta_{sol} \simeq 0.3$. The upper
bound $ \sin\theta_{reac} \lesssim 0.2 $ was obtained from the
non-observation of the disappearance of $\overline{\nu_e}$ in the
Chooz experiment\cite{chooz}. The masses of charged leptons are
the most precisely measured parameters of the fermions. The data
reads $m_e=0.51 MeV, m_\mu=106 MeV, m_\tau=1780 MeV$. Meanwhile,
as for neutrinos, the existing data just show that the neutrino
mass squared differences which induce the solar and atmospheric
neutrino oscillations are $\Delta m_{sol}^2 \simeq \left(
7^{+10}_{-2} \right) \times 10^{-5}~\mbox{eV}^2$ and $\Delta
m^2_{atm}\simeq \left( 2.5^{+1.4}_{-0.9} \right) \times
10^{-3}~\mbox{eV}^2$, respectively. With SNO\cite{SNO} and
KamLAND\cite{Eguchi:2002dm}, data have narrowed down the possible
mass spectum of neutrinos into two types, normal hierarchy
($m_1\lesssim m_2<m_3$) and inverse hierarchy ($m_3<m_1\lesssim
m_2$). If the experimental results $\Delta m_{sol}^2=m_2^2-m_1^2$
and $\Delta m_{atm}^2=|m_3^2-m_2^2|$  are accommodated to the
masses of normal hierarchy, one can obtain the following relations
for mass ratio,
\begin{eqnarray}
    \frac{m_2}{m_3} \approx
    \sqrt{\frac{\Delta m_{sol}^2}{\Delta m_{atm}^2}}=
    \left(1.7^{+1.7}_{-.6}\right) \times 10^{-1},
\end{eqnarray}
assuming $m_1$ is strongly restricted to be smaller than $m_2$ by
the order of magnitude, rather than $m_1\lesssim m_2$.

The unitary mixing matrix is defined via
$\nu_{a}=\sum^{3}_{j=1}U_{aj}\nu_{j}~(a=e,\mu,\tau)$, where
$\nu_a$ is a flavor eigenstate and $\nu_j$ is a mass eigenstate.
Including data from SNO\cite{SNO} and KamLand\cite{Eguchi:2002dm},
the range of the magnitude of the MNS mixing matrix is given by
\cite{Maki:1962mu}\cite{Guo:2002ei},
    \bea
    |U| = \left(
        \begin{array}{ccc}
        0.79-0.86 & 0.50-0.61 & 0-0.16 \\
        0.24-0.52 & 0.44-0.69 & 0.63-0.79 \\
        0.26-0.52 & 0.47-0.71 & 0.60-0.77
        \end{array}\right)
    \label{mixnum}
    \eea
at the 90\% confidence level. It can be readily recognized that
the central values of elements in the mixing matrix in
Eq.(\ref{mixnum}) are pointing likely numbers,
$\sin\theta_{sol}=\frac{1}{\sqrt{3}}$ and
$\sin\theta_{atm}=\frac{1}{\sqrt{2}}$ \cite{tribi}. If neutrino
mixing matrix is close to the lepton mixing MNS matrix given with
the values dictated by experiments as above, the neutrino mass
matrix in that basis will imbed two leading terms as
\begin{eqnarray}
    \mathcal{O} \left(m_3\right) \left(\begin{array}{ccc}
    0 & 0 & 0 \\ 0 & 1 & 1 \\ 0 & 1 & 1
    \end{array}\right) +
        \mathcal{O} \left(m_2\right)
        \left( \begin{array}{ccc}
        1 & 1 & 1 \\
        1 & 1 & 1 \\
        1 & 1 & 1 \end{array} \right) +
            \mathcal{O} \left(m_1, s_{13} \right),
    \label{target}
\end{eqnarray}
if the type of mass spectrum is normal hierarchy. The model to be
presented will explain how those different scales of elements in
the mass matrix are derived from a certain flavor symmetry.

Froggatt-Nielsen(FN) mechanism\cite{Froggatt:1978nt} with a U(1)
flavor symmetry is a commonly used strategy to construct the model
of Yukawa interaction for fermions. However, if the FN mechanism
with U(1) is taken to build neutrino Yukawa interaction and then
the seesaw mechanism\cite{Gell-Mann:vs} completes the construction
for the light neutrinos, the mass matrix can not avoid an anarchy
type\cite{Hall:1999sn},
\begin{eqnarray}
    \mathcal{O} \left(m_3\right) \left(
        \begin{array}{ccc}
        1 & 1 & 1 \\
        1 & 1 & 1 \\
        1 & 1 & 1
        \end{array}\right),
\end{eqnarray}
since breaking continuous $U(1)$ flavor symmetry alone is capable
of generating no gaps in scales of neutrino matrix elements.
Unless the approach includes somehow particular strategy, e.g., an
accidental fine tuning, the mechanism does not give rise to an
explanation of hierarchy in masses and the two large angles and
small reactor mixing angle $\theta_{e3}$. The suppressed couplings
of $\nu_e$ for the natural smallness of $\theta_{e3}$ can be
constructed using a $U(1)$ flavor symmetry in low-energy effective
theories\cite{Ling:2002nj}. However the 1-1 element is further
suppressed relative to other couplings with $\nu_e$ so that the
desired structure in Eq.(\ref{target}) would not be obtained under
a $U(1)$ flavor symmetry. We will use an idea which is that, by
extending the $U(1)$ flavor symmetry so that it contains
additional discrete abelian $Z_n$ symmetries, it is possible to
achieve relative suppression or relative enhancement among
elements in Yukawa matrices or in neutrino mass
matrix\cite{Leurer:1992wg}\cite{Berger:2000mf}\cite{Tanimoto:1999fn}.
It will be shown that a flavor symmetry $U(1) \times Z_3 \times
Z_2$ originate the particular pattern in Eq.(\ref{target}).

The model in consideration will be constructed on the basis where
the neutrino masses involve two large lepton mixing angles. The
neutrino mass matrix is derived via the seesaw mechanism by
introducing three heavy right-handed neutrinos.

In the following section, experimental data will be embedded
inside the mass matrix, for comparison with prediction. The
elements of mass matrix appear classified scale by scale.  In
Section III, we adopt the Froggatt-Nielsen(FN) mechanism
generating different orders of couplings to establish the
structure of Yukawa matrices. We will show that Yukawa couplings
generated by breaking discrete abelian flavor symmetries,
accompanied with the seesaw mechanism, can produce the distinct
scales as in the mass matrix shown as constraint in the previous
section. The prediction $(2/3)m_2/m_3 \sim \sqrt{2}
\sin\theta_{13}$ will be presented in Section IV. One of the
significant outcomes of the model is that the small mass ratio
$m_2/m_3$ and the small mixing $U_{e3}$ are resulted from breaking
of the discrete symmetries $Z_3$ and $Z_2$. Some relevant remarks
will follow in the last section.

\section{low energy constraints}
\noindent In general, a unitary mixing matrix for 3 generations of
neutrinos is given by
    \begin{eqnarray}
    \tilde{U} &=& R\left(\theta_{23}\right)
          R\left(\theta_{13},\delta\right)
          R\left(\theta_{12}\right)P(\varphi, \varphi')
    \label{fulltrans}
    \end{eqnarray}
where $R$'s are rotations with three angles and a Dirac phase
$\delta$ and the $P=Diag\left(1,e^{i\varphi/2},e^{i\varphi'/2}
\right)$ with Majorana phases $\varphi$ and $\varphi'$ is a
diagonal phase transformation. The mass matrix of light neutrinos
is given by $M_{\nu} = \tilde{U} Diag(m_1,m_2,m_3) \tilde{U}^T$,
where $m_1,m_2,m_3$ are real positive masses of light neutrinos.
Or the Majorana phases can be embedded in the diagonal mass matrix
such that
    \begin{equation}
    M_{\nu} = U Diag(m_1,\cm_2,\cm_3) U^T,
    \label{umu}
    \end{equation}
where $U \equiv \tilde{U}P^{-1}$ and $\cm_2 \equiv
m_2e^{i\varphi}$ and $\cm_3 \equiv m_3e^{i\varphi'}$. If the
transformation angles are given such that
$s_{12}=\frac{1}{\sqrt{3}}(1+\sigma), s_{23}=\frac{1}{\sqrt{2}},
s_{13}\ll 1$, where $s_{ij}$ denotes $\sin{\theta_{ij}}$ with the
mixing angle $\theta_{ij}$ between $i$-th and $j$-th generations
and $\sigma$ is also small($\ll 1$), the matrix $M_\nu$ can be
expressed as
\begin{eqnarray}
    M_{\nu} \approx
    \frac{m_1}{6}\left(\begin{array}{ccc}
    4 & -2 & -2 \\ -2 & 1 & 1 \\ -2 & 1 & 1
    \end{array}\right) +
        \frac{\cm_2}{3}\left(\begin{array}{ccc}
        1+2\sigma & 1+\frac{1}{2}\sigma & 1+\frac{1}{2}\sigma \\
        & 1-\sigma & 1-\sigma \\
        &  & 1-\sigma
        \end{array}\right) +
            \frac{\cm_3}{2}\left( \begin{array}{ccc}
            2\vartheta^2 & -\sqrt{2}\vartheta & \sqrt{2}\vartheta \\
            & 1-\vartheta^2 & -1+\vartheta^2 \\
            &  & 1-\vartheta^2 \end{array} \right),
    \label{mass1}
\end{eqnarray}
where $\vartheta=s_{13}e^{i\delta}$ with a Dirac phase $\delta$.

The mass ratio $m_1/m_2$ is assumed to be much smaller than
$m_2/m_3=0.17$. From the range of $|U|$ in Eq.(\ref{mixnum}), it
is reasonable to estimate the upper bound of the $\sigma$ and
$s_{13}$ to be about $m_2/m_3$. Since $m_1/m_2$ is negligible in
comparison with other parameters, the dimensionless matrix
obtained by a denominator $m_3/2$ reduces to an expression with
three leading parts depending on the orders and the types of
parameters,
\begin{eqnarray}
    \frac{M_{\nu}}{m_3/2}
    &=& \left(\begin{array}{rrr}
    0 & 0 & 0 \\ 0 & 1 & -1 \\ 0 & -1 & 1
    \end{array}\right) +
        \frac{2}{3}\frac{m_2}{m_3}e^{i\tilde{\varphi}}
        \left( \begin{array}{ccc}
        1 & 1 & 1 \\
        1 & 1 & 1 \\
        1 & 1 & 1 \end{array} \right) +
            \sqrt{2}s_{13}e^{i\delta}
            \left( \begin{array}{rrr}
            0 & -1 & 1 \\
            -1 & 0 & 0 \\
            1 & 0 & 0 \end{array} \right) \nonumber \\
                &+&
                \mathcal{O} \left(\frac{m_1}{m_3}, \frac{m_2}{m_3}\sigma,
                s_{13}^2, etc. \right)
    \label{morder}
\end{eqnarray}
where $\tilde{\varphi}=\varphi-\varphi'$. The entries in the three
matrix terms represent exact ones or exact zeros while small
contributions are all collected in the last term. The third term
that represents the contribution of the small mixing angle $s_{13}
\equiv \sin{\theta_{13}}$ is taken as not smaller than the ratio
$m_1/m_3$ and other parameters mentioned in the fourth term. The
value of $m_2/m_3$ is comparable to the upper bound of
experimental value of $\sin{\theta_{13}}$. If some theoretical
approaches or future experiments indicate the common size of the
two values, the constraint of the large mixing angle $\theta_{12}$
implies that a difference between the Majorana phases and the
Dirac phase $\varphi-\varphi'-\delta$ should be about $\pi/2$ to
prevent the 1-2 and 1-3 matrix elements from the suppression by
eliminating each other in two terms in Eq.(\ref{morder}). Since
the order of the third term with $s_{13}$ cannot exceed that of
the second term with $m_2/m_3$ in any circumstance, the matrix
$M_\nu /m_3$ consists of mainly two scales of elements,
$\mathcal{O}(1)$ and $\mathcal{O}(m_2/m_3)$.

\section{A model with $U(1) \times Z_3 \times Z_2$}

\noindent The Yukawa interaction of leptons and the lepton number
violating mass term of right-handed neutrinos
\begin{eqnarray}
    - \mathcal{L} = H \yuk_\ell L_e \bar{e}_R
    + H \yuk_\nu L_e \bar{N}_R + \frac{1}{2} M_R N_R N_R
\end{eqnarray}
consist of $3 \times 3$ matrices $\yuk_\ell, \yuk_\nu$ and $M_R$,
where there are three right-handed neutrinos. The FN
mechanism\cite{Froggatt:1978nt} is performed to generate the above
matrices hereafter. In models whose flavor symmetry contains three
distinct components $U(1) \times Z_3 \times Z_2$, we introduce
three singlet scalars, $S_0, S_1$ and $S_2$, with flavor charges
\begin{eqnarray}
    S_0(-1,0,0), \qquad S_1(0,-1,0)\;, \qquad S_2(0,0,-1)\;,
\end{eqnarray}
where the three elements in a parenthesis are three quantum
numbers of each field under $U(1), Z_3$, and $Z_2$, respectively.
Let the Higgs be neutral under the flavor symmetry henceforth. The
charges of fermions under the symmetry $U(1) \times Z_3 \times
Z_2$ are denoted as
\begin{eqnarray}
    L_e(p^l_i,q^l_i,r^l_i), \qquad
    N_R(p^r_i,q^r_i,r^r_i), \qquad
    e_R(p^e_i,q^e_i,r^e_i),
    \label{fcharges}
\end{eqnarray}
where $i$ denotes a generation. Then, the contributions to the
Yukawa matrices arise from flavor invariant terms in
\begin{eqnarray}
    L_i\overline{e}_jH
        \left ({{S_0}\over {\Lambda _L}}\right)^{p_{ij}^l}
        \left ({{S_1}\over {\Lambda _L}}\right )^{q_{ij}^l}
        \left({{S_2}\over {\Lambda _L}}\right )^{r_{ij}^l}
    +L_i\overline{N}_jH
        \left ({{S_0}\over {\Lambda _L}}\right)^{p_{ij}^n}
        \left ({{S_1}\over {\Lambda _L}}\right )^{q_{ij}^n}
        \left({{S_2}\over {\Lambda _L}}\right )^{r_{ij}^n},
\label{fnmech}
\end{eqnarray}
where the exponents to the scalar fields $S_0, S_1$, and $S_2$ are
given with sums of the charges of fermions in Eq.(\ref{fcharges}).
The sums $p_{ij}^a$, $q_{ij}^a$, and $r_{ij}^a$ are calculated as
follows\cite{Leurer:1992wg}\cite{Berger:2000mf}: A sum of $U(1)$
charges $p_{ij}^a$ is obtained straightforwardly,
$p_{ij}^l=p_i^l+p_j^e, p_{ij}^n=p_i^l+p_j^r$. On the other hand, a
sum of $Z_n$ charges $q_{ij}^a$ or $r_{ij}^a$ needs to be taken
extra care of for the property of discrete abelian symmetry. If
$q_i^l+q_j^e$ is less than 3, $q_{ij}^l=q_i^l+q_j^e$. If
$q_i^l+q_j^e$ is equal to or larger than 3,
$q_{ij}^l=q_i^l+q_j^e-3$, that is, the charges are modded out by 3
and so are  the $q_{ij}^e$ and the $q_{ij}^n$. The particular rule
for the charges of $Z_3$ symmetry is denoted by brackets
$[\quad]_3$ such as $q_{ij}^l=[q_i^l+q_j^e]_3$ and
$q_{ij}^n=[q_i^l+q_j^r]_3$. Likewise, $r_{ij}^l=[r_i^l+r_j^e]_2$
and $r_{ij}^n=[r_i^l+r_j^r]_2$. Also, the mass matrix $M_R$ of the
flavor invariant right-handed neutrino mass term is replaced by
couplings including singlet particles,
\begin{eqnarray}
    \frac{1}{2} \Lambda_R N_R N_R
        \left ({{S_0}\over {\Lambda _L}}\right)^{p_{ij}^r}
        \left ({{S_1}\over {\Lambda _L}}\right )^{q_{ij}^r}
        \left({{S_2}\over {\Lambda _L}}\right )^{r_{ij}^r},
\label{fnright}
\end{eqnarray}
where $p_{ij}^r=p_i^r+p_j^r, q_{ij}^r=[q_i^r+q_j^r]_3$, and
$r_{ij}^r=[r_i^r+r_j^r]_2$.

The scalar fields in general can have different vacuum expectation
values $<S_0>, <S_1>$ and $<S_2>$. These can be related to a
common expansion parameter $\lambda$ by setting
\begin{eqnarray}
    {{<S_0>}\over {\Lambda _L}} \equiv \lambda,\qquad
    {{<S_1>}\over {\Lambda _L}} \equiv \lambda ^\alpha,\qquad
    {{<S_2>}\over {\Lambda _L}} \equiv \lambda ^\beta.
\end{eqnarray}
In general, one may take $\alpha \ne \beta$, but for our
particular models we assume $\alpha=\beta$. The scale $\Lambda _L$
where massive states are integrated out of the fundamental theory
to produce an effective theory, is assumed to be larger than the
vev $<S_0>$ of the singlet scalar field so that the parameter
$<S_0>/\Lambda _L$ is smaller than one and is called $\lambda$.
Then the generated terms in charged lepton Yukawa matrix will be
of order $\lambda ^{p_{ij}^l+\alpha(q_{ij}^l+r_{ij}^l)}$. We will
restrict our attention to flavor charges that are non-negative.
Even though only the standard model fields plus right neutrinos
are concerned here, the non-negativeness is motivated by analytic
superpotential whose terms carry charges $p_{ij},q_{ij},
r_{ij}\geq 0$. The lagrangian with broken flavor symmetry reduces
to
\begin{eqnarray}
    - \mathcal{L} = H L_e \bar{e}_R \lambda^{p_{ij}^l+\alpha(q_{ij}^l+r_{ij}^l)}
    + H L_e \bar{N}_R \lambda^{p_{ij}^n+\alpha(q_{ij}^n+r_{ij}^n)}
    + \frac{1}{2} \Lambda_R N_R N_R \lambda^{p_{ij}^r+\alpha(q_{ij}^r+r_{ij}^r)}
\label{lagra}
\end{eqnarray}
There are undetermined order one coefficients multiplying these
terms, and we assume that those coefficients are sufficiently
close to one so as not to influence the hierarchy, i.e. somewhat
greater than $\lambda $ and somewhat less than $1/\lambda$.

The flavor charges $(p_i, q_i, r_i)$'s of $i$-generation fields
are assigned as follows, if there are three families of
right-handed neutrinos, $n_r=3$;
\begin{eqnarray}
    \begin{array}{lllcccccc}
            & &                & & i=1     & & i=2     & & i=3 \\
        L_e & & (p^l,q^l,r^l): & & (0,2,0) & & (0,0,1) & & (0,0,1) \\
        N_R & & (p^r,q^r,r^r): & & (2,2,1) & & (0,0,0) & & (0,0,0) \\
        e_R & & (p^e,q^e,r^e): & & (3,0,0) & & (2,0,1) & & (0,0,1)
    \end{array}
\label{charges}
\end{eqnarray}
The mass matrix of right-handed neutrinos constructed by applying
the flavor charges to Eq.(\ref{lagra}) is
\begin{eqnarray}
    M_R \sim \left( \begin{array}
    {ccc}
    \lambda^{4+\alpha} & \lambda^{2+3\alpha} & \lambda^{2+3\alpha} \\
    \lambda^{2+3\alpha} & 1 & 1 \\
    \lambda^{2+3\alpha} & 1 & 1 \end{array} \right)\Lambda_R,
    \label{right}
\end{eqnarray}
and the Yukawa matrices of neutrinos and charged leptons are
\begin{eqnarray}
        \yuk_\nu \sim \left( \begin{array}
        {ccc}
        \lambda^{2+2\alpha} & \lambda^{2\alpha} & \lambda^{2\alpha} \\
        \lambda^{2+2\alpha} & \lambda^\alpha & \lambda^\alpha \\
        \lambda^{2+2\alpha} & \rho \lambda^\alpha & \lambda^\alpha \end{array}
        \right), \quad
    \yuk_\ell \sim \left( \begin{array}
    {ccc}
    \lambda^{3+2\alpha} & \lambda^{2+3\alpha} & \lambda^{3\alpha} \\
    \lambda^{3+\alpha} & \lambda^2 & 1 \\
    \lambda^{3+\alpha} & \lambda^2 & 1 \end{array} \right),
    \label{yukawa}
\end{eqnarray}
where the $\rho=1$ in $\yuk_\nu$, but later the $\rho$ will have
different magnitude other than 1 in the basis of $M_R$ diagonal.
The exponents to $\lambda$ in the above matrices are controlled by
sum rules. The $p_{ij}$ the sum of the $U(1)$ charges has a rule
among themselves, $p_{ii}+p_{jj}-p_{ij}-p_{ji}=0,$ for $i \ne j.$
Meanwhile, for the $q_{ij}$ the modulated sum of the $Z_3$
charges, $q_{ii}+q_{jj}-q_{ij}-q_{ji}$ does not always vanish but
can reduce to $\pm 3$. Likewise, $r_{ii}+r_{jj}-r_{ij}-r_{ji}$
does reduce to either $0$ or $\pm 2$ \cite{Berger:2000mf}. If one
employs just a continuous $U(1)$ flavor symmetry, the type of
neutrino mass matrix obtained via seesaw mechanism is democratic,
which is caused by a feature of $U(1)$, the vanishing sum rule.
The suppressed elements in the neutrino mass matrix can be derived
using a discrete symmetry, if the charges of fields result in
non-zero sum rule.

\section{masses of light neutrinos}
\noindent The light neutrino masses are generated via the seesaw
mechanism $M_\nu=\yuk_\nu M_R^{-1} \yuk_\nu^{T}$. Now on, the
basis is switched to the one with right-handed neutrino mass
eigenstates corresponding to mass eigenvalues $(M_1, M_2, M_3)$.
The transformation of right-handed neutrinos may give rise to a
change in the Yukawa matrix $\yuk_\nu$, which effect can appear as
the suppression by $\rho$ factor smaller than 1. Other effect than
the $\rho$ factor from the transformation of basis is nothing but
rotations with angles about $\lambda^{2+3\alpha}$ which is
negligible. Such a transformation does not change the order of
magnitude of other contributions in $\yuk_\nu$. The light neutrino
mass is obtained as
\begin{eqnarray}
    M_\nu
    & \sim &
    \frac{v^2 \lambda^{2\alpha}}{M_3}\left(\begin{array}
    {ccc}
    \lambda^{2\alpha} & \lambda^{\alpha} & \lambda^{\alpha} \\
    \lambda^{\alpha} & 1 & 1 \\
    \lambda^{\alpha} & 1 & 1
    \end{array}\right) +
        \frac{v^2 \lambda^{2\alpha}}{M_2}\left(\begin{array}
        {ccc}
        \lambda^{2\alpha} & \lambda^\alpha & \lambda^\alpha \\
        \lambda^\alpha & 1 & \rho \\
        \lambda^\alpha & \rho & \rho^2
        \end{array}\right) +
            \frac{v^2 \lambda^{4+4\alpha}}{M_1}\left(\begin{array}
            {ccc}
            1 & 1 & 1 \\
            1 & 1 & 1 \\
            1 & 1 & 1
            \end{array}\right),
    \label{lightm1}
\end{eqnarray}
where $v$ is the vacuum expectation value of the doublet Higgs.

As shown in Eq.(\ref{charges}), there is no distinguishability
between the second and the third neutrinos under the flavor
symmetry. The right-handed neutrinos described in Eq.(\ref{right})
may have two degenerate masses, $M_2=M_3$, or may have three
different masses with maximal 2-3 mixing angle. If $M_2<M_3$, the
second term in Eq.(\ref{lightm1}) will exceed the first term,
which is not the case to be concerned in this model. The leading
orders in the elements of neutrino mass matrix can be arranged in
three parts, with $M_1 \sim \lambda^{4+\alpha}\Lambda_R$ and $M_2
\sim M_3 \sim \Lambda_R$,
\begin{eqnarray}
    M_\nu & \sim &
    \frac{v^2 \lambda^{2\alpha}}{\Lambda_R} \left\{
        \left(\begin{array}{ccc}
        0 & 0 & 0 \\
        0 & 1 & 1 \\
        0 & 1 & 1
        \end{array}\right) +
            \lambda^\alpha\left(\begin{array}{ccc}
            1 & 1 & 1 \\
            1 & 1 & 1 \\
            1 & 1 & 1
            \end{array}\right) +
                \lambda^\alpha\left(\begin{array}{ccc}
                0 & 1 & 1 \\
                1 & 0 & 0 \\
                1 & 0 & 0
                \end{array}\right)
                + \mathcal{O}(\lambda^{2\alpha}) \right\},
    \label{lightm}
\end{eqnarray}
which shows that the suppression due to the change of the basis
does not affect the order of those three terms. Thus, the masses
of the light neutrinos are predicted only in terms of the scale of
lepton number violation $\Lambda_R$ and the breaking scale of the
discrete flavor symmetry $Z_3 \times Z_2$,
$\lambda^\alpha\equiv<S>/M_\Lambda$. The breaking scale of
discrete symmetries may be the same as that of the $U(1)$ flavor
symmetry, $\alpha=1$, or may not. But, before removing $\alpha$
which denoted the contribution of discrete symmetry with
$\lambda^\alpha$, it is worth stressing that the particular
structure of the matrix in Eq.(\ref{lightm}) was able to be
derived only due to the discrete symmetries. If $\alpha=1$, the
model predicts in comparison of Eq.(\ref{morder}) and
Eq.(\ref{lightm})
\begin{eqnarray}
    && \frac{2}{3} \frac{m_2}{m_3} \sim
            \sqrt{2} \sin\theta_{13} \sim\lambda,
    \label{predict}
\end{eqnarray}
and for charged leptons with the Yukawa matrix in
Eq.(\ref{yukawa})
\begin{eqnarray}
    && \frac{m_e}{m_\tau} \sim \lambda^5,\qquad
    \frac{m_\mu}{m_\tau} \sim \lambda^2,
    \label{predictlep}
\end{eqnarray}
which are consistent to each other as well as experiments, while
they are derived in terms of a single parameter $\lambda$ induced
from the discrete symmetry breaking. Two relations in
Eq.(\ref{predictlep}) give rise to estimation of $\lambda$ about
0.19 and about 0.24, respectively. They would provide an
appropriate range for values in Eq.(\ref{predict}).

\section{Remarks}

\noindent It was shown that a $U(1) \times Z_3 \times Z_2$ flavor
symmetry invariant model of Yukawa interaction gives rise to the
prediction of Eq.(\ref{predict}) and Eq.(\ref{predictlep}) through
a series of mechanisms of breaking the symmetry and the seesaw.
The above prediction is obtained with Yukawa matrices at lepton
number violating scale $M_1$ or $M_3(\Lambda_R)$. The test of
$\sin\theta_{13}$ with future experiments requires a detailed
model that involves the choice of order one coefficients and
running of Yukawa couplings. With respect to RG running, the mass
ratios mentioned in Eq.(\ref{predict}) and in
Eq.(\ref{predictlep}) are stable in Standard Model or in Minimal
Sypersymmetric  Standard Model (MSSM)
\cite{Fusaoka:1998vc}\cite{Koide:2000ks}. However the running of a
neutrino mixing angle like $\theta_{13}$ in Eq.(\ref{predict})
reveals different aspects depending on models. It was pointed out
in Ref.\cite{Antusch:2003kp} that the RG evolution of
$\theta_{13}$ may be positive or negative relying on whether a
model in MSSM includes CP phases or not.

One can recognize that the model obtains the large mixing between
the second and the third generations simultaneously both in the
Yukawa matrix of charged leptons $\yuk_\ell$ in Eq.(\ref{yukawa})
and in the neutrino mass matrix. The MNS mixing matrix, the
multiplication of the transposed mixing matrix of charged leptons
and the mixing matrix of neutrinos\cite{Maki:1962mu}, can maintain
still large or maximal mixing due to imaginary phases even if both
mixing matrices have large or maximal angles for the relevant
mixing before the multiplication.

Under the flavor symmetry, only the first generation of neutrinos
can be distinguished from others. Other two generations of
neutrinos cannot be distinguished from each other in terms of
their charges no matter which are left-handed or right-handed,
although three generations of the left-handed are classified
within weak interaction. With the development to this point, there
is no way to state that the right-handed are of two generations or
they are of three generations. Some phenomenology with righthanded
neutrinos, e.g., leptogenesis, may help further setting of the
model.

In the model derived by FN mechanism extended with additional
discrete symmetries, it was shown that the small but moderate
neutrino mass ratio $m_2/m_3$ is a consequence from breaking the
discrete symmetries. The small value for $\theta_{13}$ is also
explained as another consequence of discrete symmetry breaking,
which is not plausible only by a continuous U(1) symmetry.

\begin{acknowledgments}
The author thanks Sin Kyu Kang for helpful comments.
\end{acknowledgments}

\end{document}